\documentclass{jltp}

\usepackage{graphicx} 

\title{Self--Consistent Conserving Theory for Quantum Impurity Systems:\\
Renormalization Group 
Analysis\thanks{Dedicated to 
Peter W\"olfle on the occasion of his 60th birthday.}}

\author{Stefan Kirchner and Johann Kroha}

\address{
Institut f\"ur Theorie der Kondensierten Materie, 
Universit\"at Karlsruhe\\
Postfach 6980,
76128 Karlsruhe,
Germany}

\runninghead{S. Kirchner and J. Kroha}{
Self--Consistent Conserving Theory for Quantum Impurity Systems}

\begin{document}

\maketitle

\begin{abstract}

We review the diagrammatic, conserving theory for quantum impurities
with strong on-site repulsion.
The method is based on  auxiliary particle technique, where Wick's
theorem is valid, which opens up the possibility for
generalizations to more complicated
situations. An analysis in terms of the
perturbative renormalization group (RG) shows that on the level of the
Conserving T--matrix Approximation the theory correctly describes 
the RG flow, including the non-scaling of potential scattering 
terms and the correct Kondo temperature.

\end{abstract}

\section{INTRODUCTION}
\label{sec1}

The physics of correlated electron systems on a lattice,
such as heavy fermion compounds or narrow band metals is determined by
a strong on--site repulsion between the electrons. It can lead to the 
formation of local magnetic moments and, subsequently, either to the Kondo
effect\cite{hewson.93} 
or to magnetic ordering, to a Mott--Hubbard metal--insulator
transition, and even to superconductivity. 
The theoretical description of such systems has been greatly advanced
by the dynamical mean field theory (DMFT),\cite{metzner.89,kotliar.96}
where the lattice system is mapped onto an effective single--impurity 
Anderson model, embedded
self--consistently in a correlated electron bath. Thus, the complexity 
of the lattice system is reduced to that of a quantum impurity problem.
Quantum impurity systems are also of central interest in their own 
right: The Anderson impurity model is the standard model, e.g., for the
description of quantum dots in the Coulomb blockade and Kondo 
regimes\cite{glazman.88,lee.88,schoen.97}, 
of scanning tunneling spectroscopy (STS) on
magnetic ions on a metal surface\cite{berndt.98,crommie.98,ujsaghy.00}, 
and for the investigation of
single--impurity physics in heavy fermion systems\cite{baer.97,reinert.01}.
Therefore, it is highly desirable to construct theoretical methods
flexible enough to describe quantum impurities with a complex 
structure of the fermionic bath (in DMFT), with multiple local
orbitals (in rare earth systems, quantum dots etc.), and which can
readily be generalized to non--equilibrium transport. Existing exact
solution methods like Bethe ansatz, conformal field theory (CFT), and
numerical renormalization group (NRG) are lacking this flexibility 
to a large extent. 

A general method, 
based on the auxilary particle technique\cite{barnes.76} (where
Wick's theorem is valid), which is 
capable of describing quantum impurity systems over the complete
temperature range from the weak to the strong coupling regime
and which, as a diagrammatic method, 
offers the above--mentioned flexibility, has
been envisaged by Peter W\"olfle starting in the late 1980´s, even before
the advent of DMFT and before the technological break--through that
made the fabrication of Kondo 
quantum dots\cite{goldhaber.98,kouwenhoven.98,weis.98} possible.
Such a method, now termed {\it Conserving T--Matrix Approximation}
(CTMA), has subsequently been 
developed\cite{kroha.92,kroha.97,schauerte.00,haule.01}, and has been shown to
describe, as an example of physical quantities, the spin susceptibility
of the single-- and the two--channel Anderson impurity model correctly
from the high temperature regime down to the lowest temperatures $T$
considered of about 1/100 of the Kondo 
scale $T_K$ \cite{kroha.99,kroha.01}.

The choice of diagrams comprising the CTMA has been justified in terms
of principal diagrams\cite{kroha.97,kroha.01}, where in each order of the
impurity hybridization the dominant term in the spin as well as in the
charge fluctuation channel are taken into account.
In the present paper we present another check for the validity of the CTMA:
The renormalization group flow of the coupling constants appearing in the
spin and charge vertices defined by the CTMA should be the same as the
{\it exact} flow of the respective coupling constants of the 
Anderson impurity model at least in the weak coupling regime. 
By self--consistency, the CTMA may then be expected to reach the 
correct strong coupling behavior. We note that this scheme might be extended
to be used as a guiding principle for the construction of approximative
strong coupling methods for other models like the Hubbard 
and $t$-$J$--models\cite{privcomm}. 

The paper is organized as follows.
After a brief overview of the auxiliary
particle technique and its exact properties in Section 2, 
the conserving approximations for the Anderson impurity model, the 
{\it Non--Crossing Approximation} (NCA) and the 
{\it Conserving T--Matrix Approximation} (CTMA), will be defined
in Section 3.
Section 4 contains the perturbative renormalization group analysis
of both the NCA and the CTMA in the sense discussed above. Some conlusions are
drawn in Section 5.

\section{EXACT PROPERTIES OF THE AUXILIARY PARTICLE TECHNIQUE}
\label{sec2}

The large on--site repulsion $U$ between electrons in the local  
$d$--orbital of an Anderson impurity effectively
restricts the dynamics to the sector of Fock space with
no double occupancy. It can be implemented using the auxiliary or 
slave boson method\cite{barnes.76}, where in the limit $U\to\infty$ the
creation operator for an electron with spin $\sigma$ in 
the $d$-level is written in terms of the auxiliary
fermion operators $f_{\sigma}$ and boson operators $b$ as 
$d^{\dag}_{\sigma} = f^{\dag}_{\sigma}b$.
This representation is exact, if 
the constraint on the total auxiliary particle number operator,
$\hat Q = \sum_{\sigma} f_{\sigma}^{\dag}f_{\sigma}^{\phantom{\dag}}+
b^{\dag}b \equiv 1 $, is fulfilled.
$f^{\dag}_{\sigma}$ and $b^{\dag}$ may be envisaged as creating the three
allowed states of the impurity: singly occupied with spin $\sigma$ or empty.
The Hamiltonian of a single Anderson impurity with local
level $E _d$, embedded in a sea of
conduction electrons (creation operators $c^{\dag}_{\vec k\sigma}$,
dispersion $\varepsilon _{\vec k}$, half band width $D$,
and density of states at the 
Fermi level $N_0$) via a hybridization matrix 
element $V$ then reads ($U\to \infty$),
\begin{eqnarray}
H=\sum _{\vec k,\sigma}\varepsilon _{\vec k}
c_{\vec k\sigma}^{\dag}c_{\vec k\sigma}^{\phantom{\dag}}+
E_d\sum _{\sigma} f_{\sigma}^{\dag}f_{\sigma}^{\phantom{\dag}}+
V\sum _{\vec k,\sigma}(c_{\vec 
k\sigma}^{\dag}b^{\dag}f_{\sigma}^{\phantom{\dag}} +h.c.)\ .
\label{sbhamilton}
\end{eqnarray}
In the Kondo regime, $N_0J_0\equiv N_0V^2/|E_d| \ll 1$, the model
has a low temperature scale, the Kondo temperature 
$T_K \simeq D {\rm e}^{-1/(2N_0J_0)}$, at which the system crosses over 
from singular spin scattering to a spin--screened, strong coupling
Fermi liquid (FL) ground state.\cite{hewson.93}

\subsection{Exact Projection onto the Physical Fock Space}
The auxiliary particle Hamiltonian 
(\ref{sbhamilton}) is invariant under simultaneous, local $U(1)$ gauge 
transformations, $f_{\sigma}\rightarrow f_{\sigma} {\rm e}^{i\phi 
(\tau )}$, $b\rightarrow b {\rm e}^{i\phi (\tau )}$, with 
$\phi (\tau )$ an arbitrary, time dependent phase. 
While the gauge symmetry guarantees the conservation of the local,
integer charge $Q$, it does not single out any particular $Q$ sector, 
like $Q=1$.
In order to project onto the $Q=1$ sector of Fock space, 
one may use the following procedure\cite{abrikosov.65,coleman.84}:  
Consider first the grand-canonical ensemble with respect to $Q$ and
the associated chemical potential $-\lambda$.
The expectation value in the $Q=1$ subspace of any
physical operator $\hat A$ acting on the impurity states is then
obtained exactly as 
\begin{equation}
\langle \hat A\rangle =
\lim _{\lambda \rightarrow \infty}
\frac {\frac{\partial }{\partial \zeta} \mbox{tr}
       \bigl[\hat A e^{-\beta (H+\lambda Q)} \bigr] _G} 
      {\frac{\partial }{\partial \zeta} \mbox{tr}
       \bigl[ e^{-\beta (H+\lambda Q)} \bigr] _G} =
\lim _{\lambda \rightarrow\infty}\frac{\langle \hat A\rangle _G}
{\langle Q \rangle _G}\ ,
\label{projection}  
\end{equation}
where the index $G$ denotes the grand canonical ensemble, 
$\zeta$ denotes the fugacity $\zeta = {\rm e}^{-\beta\lambda}$,
and $-\lambda$ is by construction the chemical potential associated
with the local charge $Q$.
In the second equality of Eq. (\ref{projection})  
we have used the fact that any physical operator
$\hat A$ acting on the impurity states is composed of the impurity electron 
operators $d_{\sigma}$, $d^{\dag}_{\sigma}$, and thus annihilates 
the states in the $Q=0$ sector, $\hat A|Q=0\rangle =0$. 
It is obvious that the grand-canonical expectation values
involved in Eq. (\ref{projection}) may be factorized into auxiliary
particle propagators using Wick's theorem, thus allowing for
the application of standard diagrammatic techniques. For a detailed
review see Ref.\ \onlinecite{kroha.01,kroha.98}.

It is important to note that, in general, $\lambda$ plays the role of a 
time dependent gauge field. In Eq. (\ref{projection}) a time independent
gauge for $\lambda$ has been chosen. 
In this way, the projection is only performed at
one instant of time, explicitly exploiting the conservation of the local
charge $Q$. This means that in the
subsequent development of the theory, the $Q$ conservation must be 
implemented exactly. It is achieved in a systematic way by means of
conserving approximations\cite{kadanoff.61}, i.e. by deriving all
self-energies and vertices by functional derivation from one common 
Luttinger-Ward functional $\Phi$ of the fully renormalized 
Green's functions, 
\begin{equation}
\Sigma_{b,f,c} = \delta \Phi \{G_b,G_f,G_c\} /\delta G_{b,f,c}.
\label{fderiv}
\end{equation}
This amounts to calculating all quantities of the theory
in a self-consistent way, but has the great advantage that gauge 
field fluctuations need not be considered. 

\subsection{Infrared Threshold Behavior of Auxilary Propagators}
As seen from  Eq. (\ref{projection}), the limit $\lambda \rightarrow \infty$
effecting the projection onto the physical subspace
implies that the traces involved in the time ordered pseudo\-fermion 
and slave boson Green's functions $G_f$, $G_b$ 
are extended purely over
the $Q=0$ sector of Fock space, and thus the backward-in-time
contribution to the auxiliary particle propagators vanishes. 
Consequently, the auxiliary particle
propagators are formally identical to the core hole propagators
appearing in the well-known X-ray problem\cite{nozieres.69}, and the
long-time behavior of $G_f$ ($G_b$) is determined by the 
orthogonality catastrophe\cite{anderson.67} 
of the overlap of the Fermi sea without 
impurity ($Q=0$) and the fully interacting 
conduction electron sea in the presence  
of a pseudofermion (slave boson) ($Q=1$).  
It may be shown that the auxiliary particle
spectral functions have threshold behavior 
with vanishing spectral weight at $T=0$ for energies $\omega $ 
below a threshold $E_o$, and power law behavior above $E_o$,  
$A_{f,b}(\omega ) \propto \Theta (\omega - E_o ) \omega ^{-\alpha _{f,b}}$.

In the present paper we will consider only the single-channel Anderson model
with its spin screened FL ground state, leaving the impurity as a
pure potential scatterer. 
In this case, the exact threshold exponents 
may be deduced \cite{mengemuha.88,kroha.97,kroha.01}
from an analysis in terms of scattering phase shifts\cite{schotte.69}, 
using the Friedel sum rule. One obtains for spin degeneracy $N\geq 1$
and conduction channel degeneracy $M=1$, 
\begin{eqnarray}
\alpha_f = \frac{2n_d - n_d^2}{N}\ ,\qquad
\alpha_b = 1-\frac{n_d^2}{N}\, \qquad\quad (N\geq 1, M=1)\ .
\label{alpha_fb1}
\end{eqnarray}
These results have been confirmed by numerical renormalization 
group (NRG) calculations\cite{costi.94} and 
by use of the Bethe ansatz solution in connection with
boundary CFT\cite{fujimoto.96}.
We note in passing that, on the contrary, 
in the non--Fermi liquid (NFL) case of the multi-channel Kondo model
($M\geq N\geq 2$) the threshold exponents have 
been deduced by a CFT solution\cite{affleck.91} as
\begin{eqnarray}
\alpha_f = \frac{M}{M+N}\ , \qquad
\alpha_b = \frac{N}{M+N}\, \qquad\quad (M\geq N\geq 2)\ . 
\label{alpha_fb2}
\end{eqnarray}
Since the dependence of $\alpha _f$, $\alpha _b$ on the impurity occupation 
number $n_d$ shown above originates from pure potential scattering,
it is characteristic for 
the FL case. The auxiliary particle threshold exponents are, therefore,
indicators for FL or NFL behavior in quantum impurity models of the
Anderson type.

\section{CONSERVING APPROXIMATIONS}
\label{sec3}

The task in constructing a self--consistent, conserving theory for
a given system consists in finding 
the proper Luttinger--Ward functional that yields 
a correct description for the respective strongly correlated model.
In this section we describe two approximation schemes for the 
Anderson impurity model, the NCA and the CTMA.  

In Matsubara representation the
auxiliary particle Green's functions $G_{f,b}$ read in terms of the 
self--energies $\Sigma_{f,b}$,
\begin{equation}
{G}_{f,b}(i\omega_n) = \Big\{[{G}_{f,b}^0(i\omega_n)]^{-1} -
\Sigma_{f,b}(i\omega_n)\Big\}^{-1} \ ,
\label{green}
\end{equation}
where
\begin{eqnarray}
{G}_{f\sigma}^0(i\omega_n) = (i\omega_n - E_d - \lambda)^{-1}\ ,\qquad
{G}_{b}^0(i\omega_n) = (i\omega_n - \lambda)^{-1} 
\label{green00}
\end{eqnarray}
are the respective free propagators. 
Since, as a consequence of the projection 
procedure $\lambda \rightarrow \infty$, the energy eigenvalues 
of $H + \lambda Q$  scale to infinity  $\propto \lambda Q$, 
it is useful to take a gauge
$\lambda=\lambda_0+\lambda '$ with $\lambda ' \to \infty$,
to shift the zero of the auxiliary particle
frequency scale by $\lambda '$, and to determine $\lambda _0$
such that for $T=0$ the threshold of the auxiliary particle 
spectral functions lies at the frequency $\omega =0$.\cite{costi.96}
In particular, for vanishing hybridization the Green's functions 
read in this gauge,
\begin{eqnarray}
{G}_{f\sigma}^0(i\omega_n) = (i\omega_n )^{-1}\ , \qquad
{G}_{b}^0(i\omega_n) = (i\omega_n + E_d)^{-1} \ . 
\label{green0}
\end{eqnarray}

\subsection{Non--Crossing Approximation (NCA)}

\begin{figure}
\centerline{\includegraphics[width=0.8\linewidth]{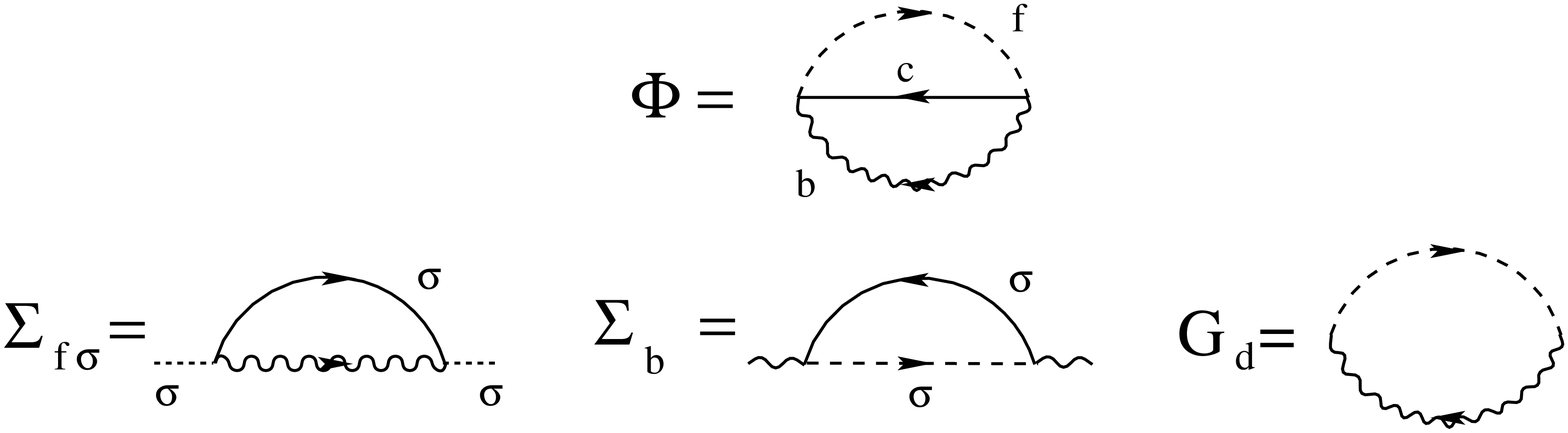}}
\caption{Diagrammatic representation of the generating functional $\Phi $ 
of the NCA. Also shown are the pseudoparticle self--energies and the
local $d$-electron Green's function derived from $\Phi$, Eqs.~(11)--(13).
Throughout this article, dashed, wavy and solid lines represent fermion, 
boson, and conduction electron lines, respectively. 
}
\label{fig1}
\end{figure}

The NCA as the simplest conserving approximation for the
Anderson impurity model is motivated as a self--consistent
expansion in terms of the hybridization parameter $V$
which is assumed to be small.  
The NCA generating functional $\Phi$ is shown in Fig.\ \ref{fig1}.
Also shown are the self--energies $\Sigma_{f,b}$ and the physical
Green's function for an electron in the local $d$-level, $G_d$,
generated from this by functional differentiation. They 
obey after analytic continuation to real frequencies 
($i\omega \rightarrow \omega -i0$) and projection onto the 
physical subspace the following 
equations of self--consistent second order perturbation theory 
\begin{eqnarray}
\Sigma_{f\sigma}^{(NCA)}(\omega-i0)&=&\Gamma\int 
              \frac{{d}\varepsilon}{\pi}\,
               f(\varepsilon )
              A_{c\sigma}^0(-\varepsilon)G_{b}(\omega +\varepsilon -i0)
              \label{sigfNCA}\\
\Sigma_{b}^{(NCA)}(\omega -i0)&=&\Gamma\sum _{\sigma}\int 
              \frac{{d}\varepsilon}{\pi}\,
              f(\varepsilon )A_{c\sigma}^0(\varepsilon)
              G_{f\sigma}(\omega +\varepsilon -i0)
              \label{sigbNCA}\\
G_{d\sigma}^{(NCA)}(\omega -i0)
         &=& \int  {d}\varepsilon\,  {\rm e}^{-\beta\varepsilon}
         [ G_{f\sigma}(\omega +\varepsilon -i0)A_{b}(\varepsilon )
          \nonumber\\
         &\ &\hspace*{1.8cm}-A_{f\sigma}(\varepsilon )
                   G_{b}(\varepsilon -\omega +i0) ] \ ,
          \label{gdNCA}
\end{eqnarray}
where $\Gamma = \pi N_0 V^2$, and $A_{c\sigma}^0=\frac{1}{\pi}\, 
{\rm Im}G_{c\sigma}^0/N_0$ is the
(unrenormalized) conduction electron density of states per spin,
normalized to the density of states at the Fermi level $N_0$, 
$A_{f,b}(\omega) ={\rm Im} G_{f,b}(\omega-i0)$
denote the imaginary parts of the advanced propagators, and 
$f(\varepsilon )=1/({\rm exp}(\beta\varepsilon )+1)$ is the 
Fermi distribution function. Together with the expressions 
(\ref{green}), (\ref{green0}) for the Green's functions,
Eqs. (\ref{sigfNCA})--(\ref{gdNCA}) form a set of self--consistent 
equations for $\Sigma_{b,f,c}$, comprised of all diagrams without 
any crossing propagator lines\cite{keiter.71,kuramoto.83}.
For an efficient procedure for the numerical evaluation of the 
self--consistent equations at low $T$ see Ref.\ \onlinecite{costi.96}.
The solutions of the NCA equations have threshold power law
behavior, with the exponents given by Eq. (\ref{alpha_fb2}),\cite{muha.84}
which are in disagreement with the correct exponents for the FL case,
Eq.\ (\ref{alpha_fb1}). As a consequence, the NCA does not correctly
describe the FL regime of the Anderson impurity model, producing a 
spurious low energy singularity in the local $d$-electron
spectrum, although it gives qualitatively correct results at high 
$T$ and down to temperatures of about the Kondo scale $T_K$.   

\subsection{Conserving T-Matrix Approximation (CTMA)}

\begin{figure}
\centerline{\includegraphics[width=0.8\linewidth]{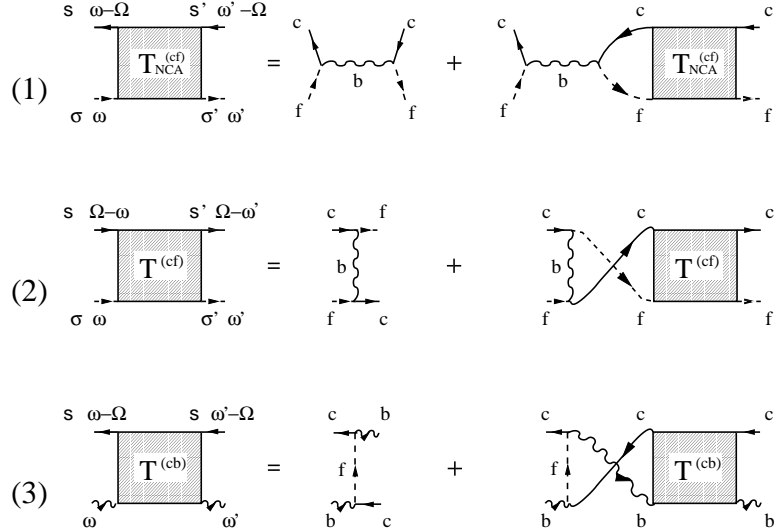}}
\caption{
Diagrammatic representation of the Bethe--Salpeter equation for (1) 
the conduction electron--pseudofermion p--h
T-matrix $T^{(cf)}_{NCA}$,  
(2) the conduction electron--pseudofermion p--p
T-matrix $T^{(cf)}$, Eq.~(\ref{cftmateq}), and (3) 
the conduction electron--slave boson 
T--matrix $T^{(cb)}$, Eq.~(\ref{cbtmateq}). The external lines 
are drawn for clarity and do not belong to the T--matrices.
}
\label{fig2}
\end{figure}
\begin{figure}
\centerline{\includegraphics[width=0.8\linewidth]{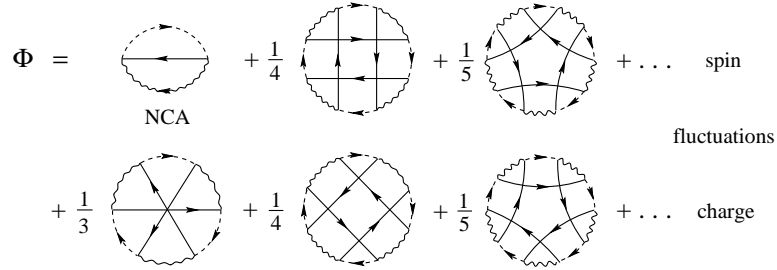}}
\caption{
Diagrammatic representation of the
Luttinger-Ward functional generating the CTMA. The terms with the conduction 
electron lines running clockwise (labelled ``spin fluctuations'') generate 
$T^{(cf)}$, while the terms with the conduction electron 
lines running counter-clockwise (labelled ``charge fluctuations'')
generate $T^{(cb)}$. The two-loop diagram is excluded,
because it is not a skeleton.
}
\label{fig3}
\end{figure}
In order to construct an aproximation which eliminates
the shortcomings of the NCA mentioned above, 
the guiding principle has been to include those 
contributions to the vertex functions which capture the 
FL physics of the single--channel Anderson model below $T_K$, i.e.
the formation of a collective singlet bound state of the local and of the
conduction electron spins.\cite{kroha.97,kroha.01} 
The onset of a bound state is usually evidenced by a pole 
in the corresponding two--particle correlation function or
T--matrix. Hence, one expects a pole in the singlet channel of the
conduction electron--pseudofermion T--matrix $T^{(cf)}$.
Taking a single slave boson propagator as the particle--particle (p--p)
and particle--hole (p--h) irreducible $c-f$ vertex (leading order in
$V$ contribution), the equations for the $c-f$ p--h and p--p
T--matrices are shown diagrammatically in Fig.\ \ref{fig2} (1), (2).
These contributions include, at any given order in the hybridization $V$,
the maximum number of spin flips possible. Considering that in the
mixed valence regime the Anderson model is dominated equally by
spin and by charge fluctuations, one should also take the 
conduction electron--slave boson T--matrix $T^{(cb)}$ into account,
which similarly includes the maximum number of charge fluctuation processes
(see Fig.\ \ref{fig2} (3)). The (linear) Bethe--Salpeter equations for   
$T^{(cf)}$, $T^{(cb)}$ read,
\begin{eqnarray}
T^{(cf)}_{\sigma\tau,\sigma '\tau '}
(i\omega _n, i\omega _n ', i\Omega _n ) =
&+&V^2G_{b}(i\omega _n + i\omega _n ' - i\Omega _n ) 
\delta _{\sigma\tau '}\delta _{\tau \sigma '}
\nonumber\\
&-&V^2T\sum _{\omega _n''}G_{b}(i\omega _n + i\omega _n '' - 
i\Omega _n ) \times 
\label{cftmateq}\\
&&\hspace*{-0.6cm}
G_{f\sigma}(i\omega _n'') \ G^0_{c}(i\Omega _n -i\omega _n '')\ 
T^{(cf)}_{\tau \sigma,\sigma '\tau '}(i\omega _n '', i\omega _n ', 
i\Omega _n ) \nonumber
\end{eqnarray}\noindent
\begin{eqnarray}
T^{(cb)\ \sigma}(i\omega _n,i\omega _n ',i\Omega _n) =
\hspace*{-0.3cm}&+&V^2G_{f\sigma}(+i\omega _n + i\omega _n ' - i\Omega _n ) 
\nonumber\\
\hspace*{-0.3cm}&-&V^2T\sum _{\omega _n''}G_{f\sigma}(i\omega _n + i\omega _n '' - 
i\Omega _n ) \times 
\label{cbtmateq}\\
&&\hspace*{-0.6cm}
G_{b}(i\omega _n'') \ G^0_{c\sigma}(-i\omega _n ''-i\Omega _n )\ 
T^{(cb)\ \sigma}(i\omega _n '', i\omega _n ', 
i\Omega _n ). \nonumber
\end{eqnarray}\noindent
\begin{figure}
\centerline{\includegraphics[width=0.8\linewidth]{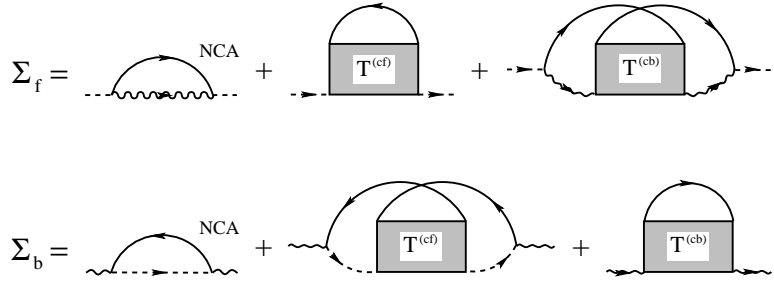}}
\caption{Diagrammatic representation of the CTMA 
pseudofermion and slave boson self--energies $\Sigma_f$, $\Sigma_b$.
}
\label{fig4} 
\end{figure}
Inserting the NCA solutions for $G_{f\sigma}$, $G_b$ into
Eq.\ (\ref{cftmateq}), one finds after analytic continuation, 
as expected, a pole in the singlet channel
of $T^{(cf)}$ at a center--of--mass frequency $\Omega_0$ which scales with
the Kondo temperature, $\Omega_0\approx - T_K$.
In order to incorporate this physics into the self--consistent 
scheme, we must construct a Luttinger--Ward functinal $\Phi$ which 
includes the total vertices 
$T^{(cf)}_{NCA}$, $T^{(cf)}$, $T^{(cb)}$ on the level of the
self--energies. This functional is shown in Fig.\ \ref{fig3}
and consists of the sum of all slave particle rings where the
bare hybridization vertices are connected by conduction electron lines
in such a way that at most two other hybridization vertices are spanned.
The set of self--energy terms deduced from $\Phi$ by functional 
differentiation is shown in Fig.\ \ref{fig4}.
In the self--consistent scheme all propagators appearing in the 
self--energies are understood to be the fully renormalized ones.
In order to avoid double counting of terms, the one-rung term
in the second diagram of $\Sigma _f$ and the one- and two-rung terms
in the third  diagram of $\Sigma _f$, Fig.\ \ref{fig4} must be
excluded. An analogous exclusion of terms must be done in $\Sigma _b$.
For a detailed description see Ref.\ \onlinecite{kroha.01}.
Note that $T^{(cf)}_{NCA}$ just corresponds to a renormalization of the 
boson propagator by the NCA self--energy and, hence, is already included
on the NCA level of self--consistent approximation. 
The self--consistent set of non--linear integral equations
resulting from the definition of the self--energies Fig.\ \ref{fig4}
and the full Green's functions Eq.\ (\ref{green}) is called CTMA.
Solving the CTMA equations selfconsistently by iteration, the 
aforementioned pole is shifted to the threshold frequency $\Omega = 0$
where it merges with the continuous spectral density and thus 
renormalizes the threshold exponents. 
\begin{figure}
\centerline{\includegraphics[width=0.48\linewidth]{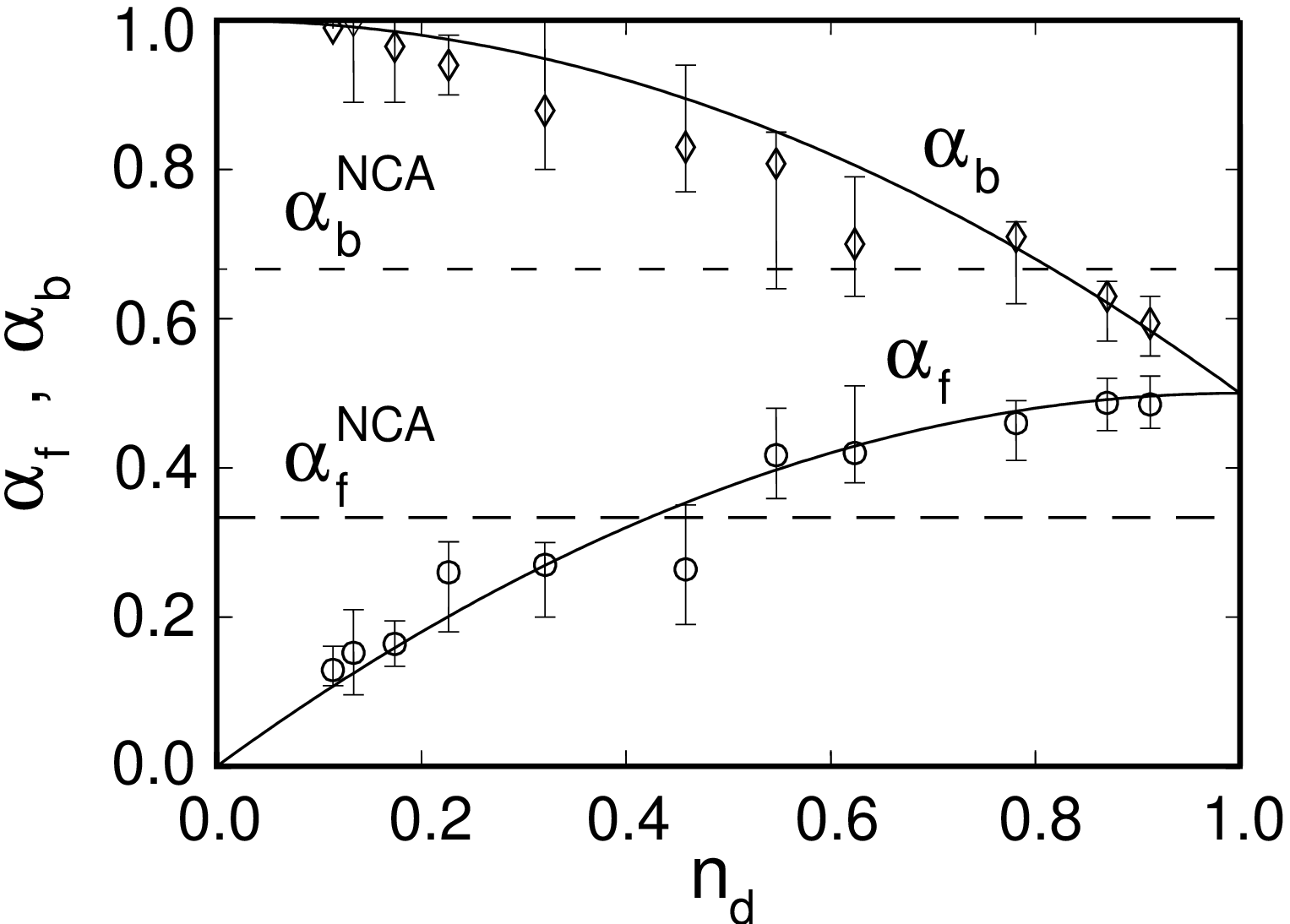}
         \hfill   \includegraphics[width=0.47\linewidth]{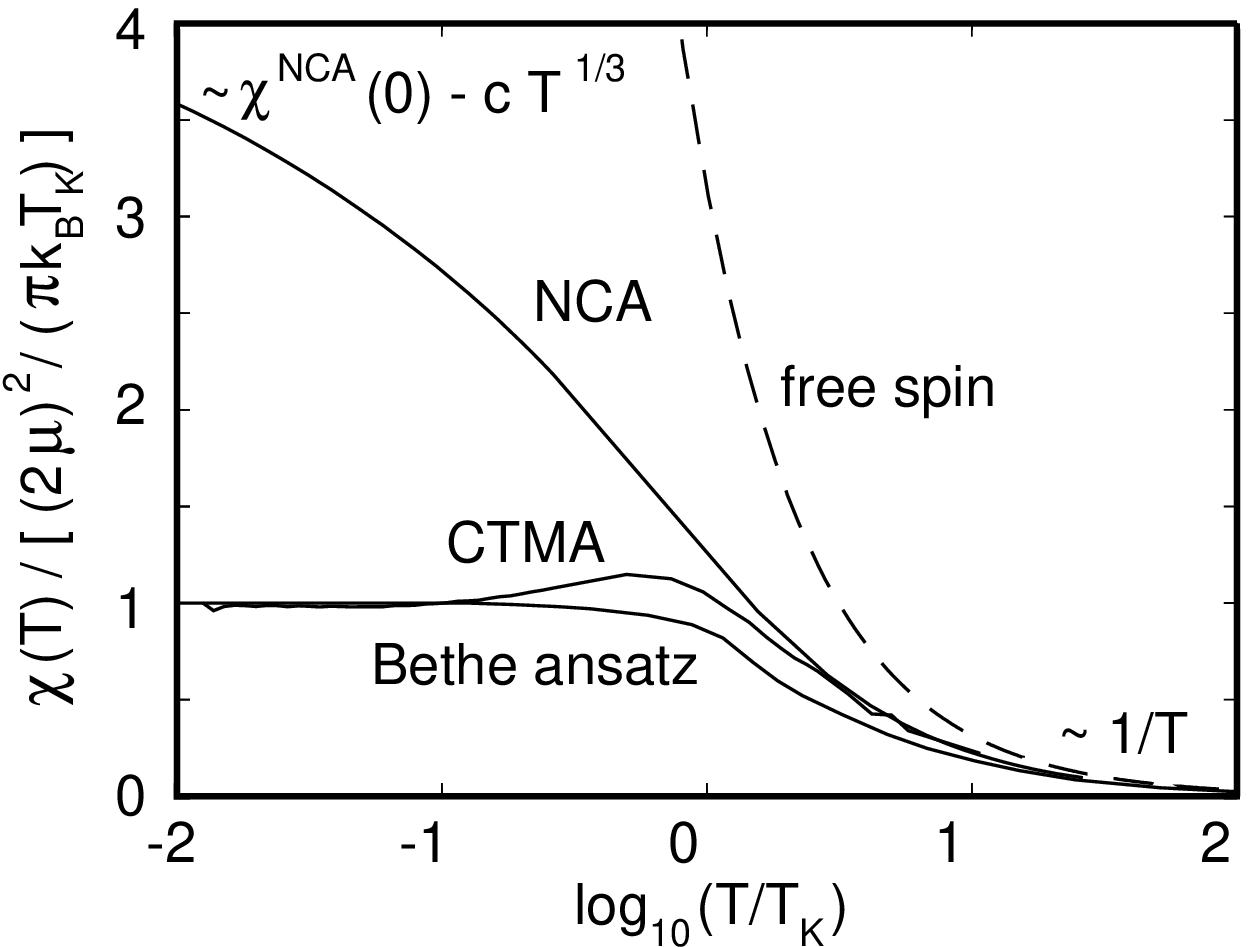}}
\caption{
Left panel:
The fermion and boson threshold exponents $\alpha _f$, $\alpha _b$
are shown for $N=2$, $M=1$ in dependence of the average impurity
occupation $n_d$. Solid lines: exact values, Eq. (\ref{alpha_fb1});
Symbols with error bars: CTMA; dashed lines: NCA.
Right panel:
Static susceptibility of the single-channel 
Anderson impurity model in the Kondo regime 
($E_d=-0.8D$, $\Gamma = 0.1D$, Land\'e factor $g=2$). 
The CTMA result is calculated using the 
The CTMA and NCA results are compared 
to the Bethe ansatz result for the Kondo model\cite{andrei.83}. 
The CTMA susceptibility obeys scaling behavior in accordance
with the exact results (not shown).
\label{fig5} 
}
\end{figure}
As signatures of the correct FL behavior below $T_K$ it was found
from the numerical solutions of the CTMA equations first that   
the correct FL exponents of the auxiliary particle propagators,
Eq.\ (\ref{alpha_fb1}), are reproduced within the error bars of the
numerical solution, and second that the static 
impurity spin susceptibility $\chi$
shows Pauli behavior ($\chi(T) \approx const.$) for $T< T_K$ down to
the lowest $T$ considered, $T\approx 10^{-2}\cdot T_K$ (see Fig.\ \ref{fig5}).

\section{PERTURBATIVE RG ANALYSIS}
\label{sec4}

Having introduced the CTMA as a method to describe quantum impurities 
in the strong coupling region, it is also important to know how it
performs in the weak to intermediate coupling region, i.e. how within 
the CTMA the coupling constants for spin and potential scattering 
are renormalized in the sense of the perturbative renormalization
group (RG). To obtain the correct weak to intermediate  coupling 
renormalization within a self--consistent treatment turns out to 
be a non--trivial task. As seen below, 
not even within the NCA, which is a self--consistent expansion in 
the small parameter $V$ (hybridization), the correct weak coupling
renormalization is produced for the Anderson impurity model.

In this section the coupling constant renormalization for 
spin and potential scattering will be analyzed within the
perturbative RG\cite{anderson.69,anderson.70}.
In doing so we will assume to be in the Kondo regime of the Anderson model,
where $E_d<0$ and $\Gamma/|E_d| \ll 1$.

In the (anisotropic) Kondo model the 
local spin--conduction electron interaction vertex is defined as
\begin{equation}
J_{\perp}(S^+\sigma ^- +S^- \sigma ^+) + J_{||} S_z\sigma _z \ .
\label{kondo_aniso}
\end{equation}
For isotropic coupling, $J_{\perp}= J_{||}=J$, this reduces to 
the regular Heisenberg coupling $J\, \vec S \cdot \vec \sigma$.
In the above expressions $\vec S =(S_x,S_y,S_z)$ denotes the
impurity spin 1/2 vector operator and $\vec \sigma =
(\sigma_x,\sigma_y,\sigma_z)$ the vector of Pauli matrices. 
$S^{\pm}$ and $\sigma^{\pm}$ are the spin raising (+) and lowering (-)
operators for the local and for the conduction electron spin,
respectively. In addition, there may be a potential scattering term
which has the structure
\begin{equation}
W\, ( {\bf 1}\, {\bf 1} )
\end{equation}
with {\bf 1} the 2$\times$2 unit matrix in impurity and 
in conduction electron spin space.

\begin{figure}
\centerline{\includegraphics[width=\linewidth]{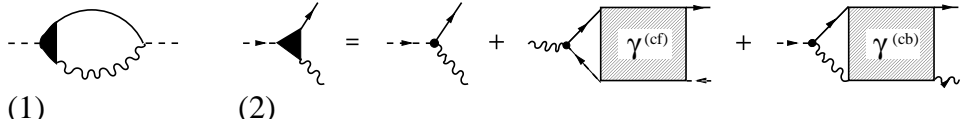}}
\caption{
(1) The fermion selfenergy expressed in terms of the exact
3--point vertex $\hat V$ (black triangle) is shown.
(2) Representation of $\hat V$ in terms of the 4--point vertices 
 $\gamma ^{cf}$, $\gamma ^{cb}$. The first diagram on the 
right--hand side is the bare 3--point vertex $V$. 
}
\label{fig6}
\end{figure}
Since in the Anderson model the hybridization $V$ (Eq.\ (\ref{sbhamilton}))
appears as the 
bare coupling constant, but not the spin and potential scattering
couplings $J_{\perp}$, $J_{||}$, and $W$, the NCA or CTMA self--energies
must first be represented in terms of the 4--point spin and potential 
vertices. This is done as follows for the pseudofermion self--energy
$\Sigma _f$. We note that 
it is not necessary to consider separately the boson self--energy
$\Sigma_b$ here, since $\Sigma _b$ will appear naturally as a 
coupling constant renormalization in the perturbative RG.
The (exact) $\Sigma _f$ can be written in terms of the exact, total
3--point hybridization vertex $\hat V$ in slave boson representation as
(see Fig.\ \ref{fig6} (1))
\begin{eqnarray}
\Sigma _{f\sigma} (\omega -i0) &=& 
V \int \frac{d\varepsilon}{2\pi} f(\varepsilon) G^0_b(\varepsilon +\omega -i0)
\times
\label{sigmafexact}\\
&\phantom{+}&\hspace*{-2cm}
\big[ \hat V (\omega -i0,\varepsilon +i0)\, G^0_{c\sigma} (-\varepsilon -i0)-
   \hat V (\omega -i0 ,\varepsilon -i0)\, G^0_{c\sigma} (-\varepsilon +i0) 
\big]
\ ,
\nonumber
\end{eqnarray}
where $G^0_b$ is the bare boson Green's function, Eq.\ (\ref{green0}).
The total 3--point vertex $\hat V$ can in turn be expressed in terms of the 
exact, total 4--point c--f and c--b vertices $\gamma ^{cf}$, $\gamma ^{cb}$,
as shown in Fig.\ \ref{fig6} (2). These 4--point vertices are comprised
of all vertex corrections connecting the f and the c line or the 
b and the c line of the 3--point vertex $\hat V$, respectively.
Note that vertex corrections connecting the f and b lines do not appear
as these are of higher order in the fugacity ${\rm e}^{-\beta\lambda}$
and thus vanish under projection onto the physical subspace.
A given approximation for the selfenergy $\Sigma _f$ (like NCA or CTMA)
corresponds to a respective approximation for   
$\gamma ^{cf}$, $\gamma ^{cb}$.

The perturbative RG analysis can now be applied to the Bethe--Salpeter
equations for the total vertices (T--matrices) 
$\gamma ^{cf}$, $\gamma ^{cb}$,
thus determining the RG flow of the spin and potential scattering
coupling constants within the given approximation.
Note that for the {\it perturbative} RG treatment all propagator
lines appearing are the bare ones,  
while in the self--consistent treatment described above 
all propagators are fully renormalized. 

\subsection{Perturbative RG for the NCA}

\begin{figure}
\centerline{\includegraphics[width=0.9\linewidth]{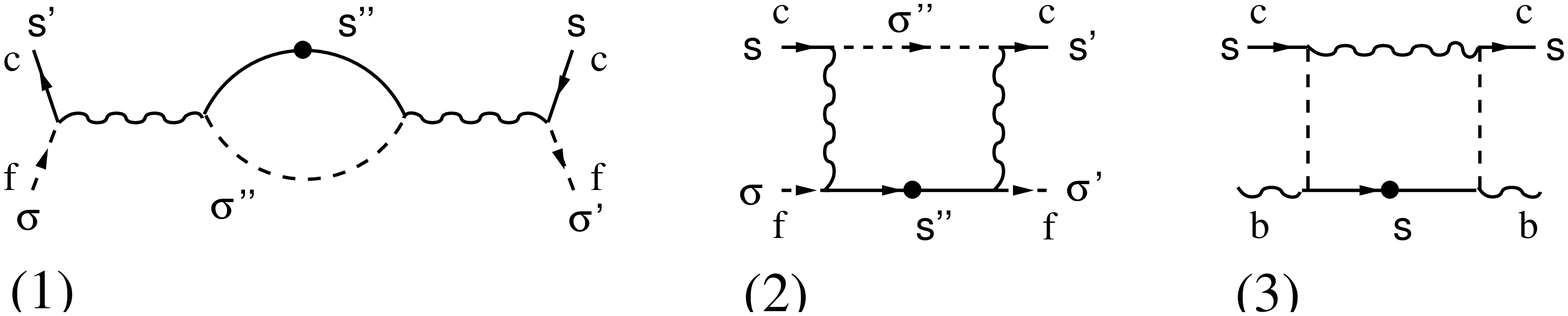}}
\caption{
perturbative RG renormalizations of the (irreducible) c--f vertex.
Diagram (1) is the NCA result, the sum of diagrams (1), (2) the
result of CTMA. (3) ist the contribution from the c--b vertex,
which is not scaling. The black dots on the conduction electron lines
indicate that the frequency integrals in those lines are 
restricted to the region $[-D,-D+dD]$.  
}
\label{fig7}
\end{figure}
Inserting the ``RPA--like'' T--matrix $T^{(cf)}_{NCA}$ for
$\gamma ^{cf}$ and the resulting $\hat V$ into Eq.\ (\ref{sigmafexact})
for the f self--energy, it is seen that the NCA is just defined by
\begin{eqnarray}\gamma ^{cf} &=& T^{(cf)}_{NCA}\\
\gamma ^{cb} &=& 0 \ ,
\end{eqnarray}
and $T^{(cf)}_{NCA}$ provides just the renormalization of the
boson propagator by the NCA self--energy, as mentiond above.
Integrating out the high energy region from the Bethe--Salpeter
equation for $T^{(cf)}_{NCA}$ and applying the usual perturbative
poor man's scaling analysis\cite{anderson.69,anderson.70},
we obtain for the renormalization of the irreducible 
conduction electron--pseudofermion vertex under a change of the
high energy cutoff from $D$ to $D-dD$ (compare Fig.\ \ref{fig7} (1))
\begin{eqnarray}
d\Lambda ^{(cf)} = - N_0 \frac{dD}{D} \sum _{s'' \sigma ''}
                    \big[ \Lambda^{(cf)} \big]_{s' \sigma '',s''\sigma}
                    \big[ \Lambda^{(cf)} \big]_{s''\sigma ' ,s  \sigma''}
\end{eqnarray}
where the spin indices $s$, $s '$, $s''$, $\sigma$, $\sigma '$, $\sigma ''$, 
are as defined in Fig.\ \ref{fig7} (1) and the (unrenormalized) 
c--f vertex is given by the inhomogeneous part of the  
Bethe--Salpeter equation (\ref{cftmateq}) [see also Fig.\ \ref{fig2} (2)],
\begin{eqnarray}
\Lambda ^{(cf)}_{s'\sigma ', s\sigma} 
         = [J_{\perp} (S^+ \sigma^- + S^- \sigma^+ ) +
            J_{||} S_z \sigma _z + W {\bf 1} {\bf 1}]_{s'\sigma ', s\sigma}\ .
\label{gamma}
\end{eqnarray}
In Eq.\ (\ref{gamma}) $s$, $s'$ denote the conduction electron spin
indices, $\sigma $, $\sigma '$ the impurity spin indices, where an
index with (without) prime represents an ingoing (outgoing) particle.
For the Anderson model we have $J_{\perp} = J_{||} = 2W = V^2/|E_d|$,
the $|E_d|^{-1}$ arising from the bare boson propagator, Eq.\ \ref{green0}, 
taken at frequencies $|\omega |\ll |E_d|$.
However, for generality we keep the anisotropic couplings 
$J_{\perp}$, $J_{||}$, $W$ in the above equations.
$d\Lambda ^{(cf)}$ may be simplified as
\begin{eqnarray}
d\Lambda ^{(cf)} = - N_0 d{\rm ln} D 
                 &\Bigl[&  (J_{||} J_{\perp} +2W J_{\perp}) 
                     [S^+ \sigma^- + S^- \sigma^+ ] \nonumber \\
                    &+&(J_{\perp}^2 + 2 W J_{||}) [S_z \sigma _z]  
                     \label{gammaNCA}\\
                    &+&(\frac{1}{2} J_{\perp}^2 + \frac{1}{4} J_{||}^2 +W^2)
                     [{\bf 1} {\bf 1}]  \Bigr] \nonumber \ .
\end{eqnarray}
Hence, within NCA we have the RG equations
\begin{eqnarray}
\frac{dJ_{\perp}} {d{\rm ln} D} &=& - N_0 (J_{||} J_{\perp} +2W J_{\perp})\\
\frac{dJ_{||}} {d{\rm ln} D} &=& - N_0 (J_{\perp}^2 + 2 W J_{||})\\
\frac{dW} {d{\rm ln} D} &=& - N_0 
(\frac{1}{2} J_{\perp}^2 + \frac{1}{4} J_{||}^2 +W^2) \ .
\label{RG}
\end{eqnarray}
For the initial conditions of the Anderson impurity case, 
$J_{\perp,0} = J_{||,0} = 2W_0 = V^2/|E_d| =J_0$,
these are easily integrated to give
\begin{eqnarray}
J(D) = \frac{J_0}{1+2N_0 J_0 {\rm ln}\frac{D}{D_0}} \ ,
\end{eqnarray}
i.e. the spin coupling constant diverges at the Kondo temperature
$T_K=D_0 {\rm e}^{-1/(2N_0J_0)}$. However, the potential scattering 
coupling is also renormalized under the RG flow. Thus, within NCA
there is a spurious divergence of the potential scattering term at 
$T_K$ as well. The fact that within NCA potential scattering is 
incorrectly treated on the same footing as spin scattering
may be traced back to be the origin why (1) within NCA the asymmetry 
of the Kondo resonance comes out too large and why (2) NCA
gives a qualitatively wrong descripting of the Kondo resonance
in a magnetic field \cite{sellier.01}.

\subsection{Perturbative RG for the CTMA}

We now consider the coupling constant renormalization under the
RG flow within CTMA. The CTMA f self--energy (Fig.\ \ref{fig4}) is
generated by inserting, in addition to the NCA contribution, the
ladder T--matrices, Eqs.\ (\ref{cftmateq}), (\ref{cbtmateq}), 
for $\gamma ^{(cf)}$, $\gamma ^{(cb)}$.
This is seen by comparing Fig.\ \ref{fig6} with Fig.\ \ref{fig4}.\cite{note}
 i.e. we
have {\it in addition} to the vertices of the previous section,
\begin{eqnarray}
\gamma ^{cf} &=& T^{(cf)}\\
\gamma ^{cb} &=& T^{(cb)}\ .
\end{eqnarray}
The resulting {\it additional} c--f 
vertex renormalization under cutoff reduction is (see Fig.\ \ref{fig7} (2)) 
\begin{eqnarray}
d\Lambda ^{(cf)} = + N_0 \frac{dD}{D} \sum _{s'' \sigma ''}
                    \big[ \Lambda^{(cf)} \big]_{s' \sigma ',s''\sigma ''}
                    \big[ \Lambda^{(cf)} \big]_{s''\sigma '' ,s \sigma} \ ,
\end{eqnarray}
which can again be simplified as
\begin{eqnarray}
d\Lambda ^{(cf)} = + N_0 d{\rm ln} D 
                 &\Bigl[&  (-J_{||} J_{\perp} +2W J_{\perp}) 
                     [S^+ \sigma^- + S^- \sigma^+ ] \nonumber \\
                    &+&(-J_{\perp}^2 + 2 W J_{||}) [S_z \sigma _z]  
                   \label{gammaCTMA}\\
                    &+&(\frac{1}{2} J_{\perp}^2 + \frac{1}{4} J_{||}^2 +W^2)
                     [{\bf 1} {\bf 1}]  \Bigr] \nonumber \ .
\end{eqnarray}

Finally, we also need to investigate the renormalization due to
$\gamma ^{(cb)}$ (Fig.\ \ref{fig7} (3)). 
This term is obviously diagonal in spin space (potential scattering).
Its amplitude is
\begin{eqnarray}
d\Lambda ^{(cb)} &=& - N_0 V^4 \int _{-D}^{-D+dD}d\varepsilon 
\times \nonumber \\
&&G^{0}_{f\sigma}(\omega +\varepsilon)  
G^{0}_{f\sigma}(\omega ' +\varepsilon) G^0_{b}(\varepsilon -\Omega) \Biggr|
_{\Omega=0, \omega,\omega ',\varepsilon\ll D <|E_d|}
\\
&=&- N_0\frac{V^4}{|E_d|} \frac{dD}{D^2} \nonumber \ ,
\end{eqnarray}
or, with $W=V^2/(2|E_d|)$,
\begin{eqnarray}
dW = - 4 N_0 |E_d| W^2 \frac{dD}{D^2} \ .
\label{no}
\end{eqnarray}
This correction is not logarithmic (non-scaling), 
and the integration of Eq.\ (\ref{no}) yields a nondivergent result. 
Hence, it need not be considered in the RG flow. 

Adding up the contributions Eqs.\ (\ref{gammaNCA}) and (\ref{gammaCTMA})
it is seen that the potential terms always cancel each other. Thus,
there is no renormalization of the potential term.
The resulting RG equations of the CTMA are
\begin{eqnarray}
\frac{dJ_{\perp}} {d{\rm ln} D} &=& - 2 N_0 J_{||} J_{\perp}\\
\frac{dJ_{||}} {d{\rm ln} D} &=& - 2 N_0 J_{\perp}^2 \\
\frac{dW} {d{\rm ln} D} &=& 0 \ .
\end{eqnarray}
These are identical to the perturbative RG equations of the 
original Kondo model. 
This proves that the CTMA incorporates the complete Kondo physics
also in the weak and intermediate coupling regime, where the
perturbative RG is valid.

\section{CONCLUSION}
\label{sec5}
We have given an overview of a diagrammatic, conserving approximation
the CTMA, designed to describe quantum impurity systems of the Anderson type
in the strong coupling regime. By means of a perturbative 
renormalization group analysis it was shown that, in contrast to an earlier 
approximation (NCA) it recovers the correct RG flow in the weak to
intermediate coupling region. In particular, it describes correctly,
that potential scattering terms do not scale in the Kondo and
Anderson problems, while the correct Kondo scale for the divergence
of the spin dependent part is recovered. Together with the fact that
in numerical solutions of the CTMA the auxiliary particle
threshold exponents and the spin susceptibility show the 
correct Fermi liquid behavior of the strong coupling regime,
this provides evidence that the CTMA correctly describes the
Anderson model both in the strong and in the weak coupling regime 
on the same footing. It has also the potential to be
generalized to describe more complicated situations and lattice
problems by use of the DMFT.

\section*{ACKNOWLEDGMENTS}

It is our pleasure to thank Peter W\"olfle who has provided continuing
support and guidance throughout the progress of the CTMA project. 
We would also like to thank A. Rosch, G. Sellier, P. Hirschfeld,
T. Kopp, K. Haule and T. Schauerte for valuable discussions.
This work was supported by DFG through SFB195.


\begin{thebibliography}{99}
\bibitem{note}
Note that the single--rung term of $T^{(cf)}$ or $T^{(cb)}$,
when inserted for $\gamma^{(cf)}$ or $\gamma^{(cb)}$, respectively, into
the 3--point verex $\hat V$, does not lead to an allowed diagram. 
However, the corresponding self--energy terms containing two
(bare) boson (fermion) lines are already included in the NCA
(see section 4.1). Also note that in order to avoid double
counting, the two-rung terms have to be subtracted from 
$T^{(cf)}$ or $T^{(cb)}$. Since these are, however, a finite number
of terms, they do not contribute to the re-scaling and can be neglected 
in the RG analysis.
\end{thebibliography}
\end{document}